\documentstyle[aps,preprint,floats,epsf,amsmath,eqsecnum]{revtex}
\begin{document}
\advance\textheight by 0.2in 
\draft

\title{Solvable model of a polymer in random media with long ranged 
disorder correlations}

\author{Yohannes Shiferaw and Yadin Y. Goldschmidt}

\address{ Department of Physics and Astronomy, \\
  University of Pittsburgh, Pittsburgh PA 15260, U.S.A. }

\date{\today} 

\maketitle

\begin{abstract}
We present an exactly solvable model of a Gaussian (flexible) 
polymer chain in a quenched random medium. This is the case when 
the random medium
obeys very long range quadratic correlations.   The model is solved 
in $d$ spatial dimensions using the replica method, and practically 
all the physical properties
of the chain can be found. In particular the difference between the 
behavior of a chain that is free to move and a chain with one end 
fixed is elucidated. The interesting finding is that a chain that 
is free to move in a quadratically correlated random potential behaves
like a free chain with $R^2 \sim L$, where $R$ is the end to end 
distance and $L$ is the length of the chain, whereas for a chain 
anchored at one end $R^2 \sim L^4$. 
The exact results are found to agree with an alternative
numerical solution in $d=1$ dimensions. The crossover from long 
ranged to short ranged correlations of the disorder is also explored. 
\end{abstract}
\pacs{PACS number(s): 05.40-a, 75.10.Nr, 36.20.Ey, 64.60.Cn}
\newpage
\section{Introduction}

The behavior of polymer chains in random media is a well studied problem
[1-6] that has applications in diverse fields.  Besides the polymers 
themselves this problem is directly related to the statistical mechanics
of a quantum particle in a random potential \cite{gold2}, the
behavior of flux lines in superconductors in the 
presence of columnar defects \cite{nelson,gold3}, and the problem of 
diffusion in a random catalytic environment \cite{nattermann}. Despite the 
volume of work that has been done on these problems there are still many 
unanswered questions. Most of the previous work (with the exception of 
directed polymers \cite{parisi,MP}) concentrated on disorder with short 
ranged correlations.
In this paper we consider a model with long ranged (quadratic) correlations 
of the random potential that can serve as a laboratory (toy model) since 
it can be solved exactly
using the replica method \cite{MPV}. Since some people are somewhat wary 
of the $n \rightarrow 0$ limit used in replica calculations, 
we also solve the model numerically in one dimension and obtain an 
excellent agreement with the analytical solution. More importantly,
the numerical solution enables us to explore the crossover from long 
ranged to short ranged correlations of the disorder and obtain a coherent
picture of the behavior of a Gaussian chain in a random medium.
 
The simplest model of a polymer chain in random media is a Gaussian 
(flexible) chain \cite{doi} in a medium of fixed random obstacles
\cite{mut}. In this paper we do not include a self-avoiding interaction.
This model can be described by the Hamiltonian
\begin{equation}
H=\int^L_0 du \left[\frac{M}{2} \left(\frac{d{\bf{R}}(u)}{ 
du}\right)^2+\frac{\mu}{2}{\bf{R}}^2(u)+V\left({\bf{R}}(u)\right)\right],
\label{Ham}
\end{equation}
where ${\bf{R}}(u)$ is the $d$ dimensional position vector of a point on the
polymer at arc-length $u\  (0 \leq u \leq L)$, and where $L$ is the 
contour length of the chain. 
The medium of random obstacles is described by a random potential $V({\bf{R}})$ 
that is taken from a Gaussian distribution that satisfies
\begin{equation}
\langle V({\bf{R}}) {\rangle}=0,\;        \; {\langle} 
V({\bf{R}})V({\bf{R}}'){\rangle}=f\left(({\bf{R}}-{\bf{R}}')^2\right).
\label{VV}
\end{equation}
The harmonic term in the Hamiltonian is included to mimic the effects of 
finite volume.  This is important to ensure that the model is well
defined, since it turns out that certain equilibrium
properties of the polymer diverge in the infinite volume
limit $(\mu \rightarrow 0)$. The function $f$ characterizes the correlations 
of the random potential, and will depend on the particular problem at hand.  
The parameter $M$ is 
inversely proportional to $\beta b^2$, where $\beta=(k_B T)^{-1}$, and where 
$b$ is the Khun bond step. 

Once we have defined the Hamiltonian for any chain configuration ${\bf{R}}(u)$,
we can write the partition sum (Green's function) for the set of 
paths of length $L$
that go from ${\bf{R}}$ to ${\bf{R}}'$ as
\begin{equation}
Z({\bf{R}},{\bf{R}}';L)=\int_{{\bf{R}}(0)={\bf{R}}}^{{\bf{R}}(L)={\bf{R}}'} 
[d{\bf{R}}(u)] {\exp}(-\beta H).
\label{Z}
\end{equation}
All the statistical properties of the polymer will depend on the partition 
sum.  For instance, we can calculate the averaged mean squared displacement of 
the far end of a polymer with one end that is fixed at the origin.  This is
a measure of the wandering of a tethered polymer immersed in a random medium.
This quantity can be written as 
\begin{equation}
\overline{{\langle} {\bf{R}}_T^2 (L){\rangle}}=\overline{\left(\frac{\int 
d{\bf{R}} {\bf{R}}^2 
Z({\bf{0}},{\bf{R}};L)}{\int d{\bf{R}} Z({\bf{0}},{\bf{R}};L)}\right)},
\label{rt2av}
\end{equation}
where the overbar stands for the average of the ratio  over the
realizations of the random potential.  This average is referred to as a
quenched average, as opposed to an annealed average, where the
numerator and denominator are averaged independently. For a polymer with 
one end fixed a typical conformation in a random medium is that of a tadpole.
The head of the polymer wanders far from the origin to find a 
region of favorable potential
and then the remaining chain settles itself in that region. This is at least
what is believed to happen when the disorder has short ranged correlations
\cite{cates,nattermann}.
On the other hand if the chain is not anchored but both ends are free to move, 
the head to tail mean squared displacement is given by 
\begin{equation}
\overline{{\langle} {\bf{R}}_F^2 (L){\rangle}}=\overline{\left(\frac
{\int d{\bf{R}} d{\bf{R'}}({\bf R}-{\bf R'})^2 
Z({\bf{R}},{\bf{R'}};L)}{\int d{\bf{R}} d{\bf{R'}}Z({\bf{R}},{\bf{R'}};L)}\right)}.
\label{rf2av}
\end{equation}  
In this case the chain can move as a whole to find a favorable environment
in the random medium.

In order to compute the quenched average over the random potential we apply 
the replica method.  We first introduce $n$-copies of the system and average 
over the random potential to get
\begin{equation}
Z_n(\{{\bf R}_a\},\{{\bf R}'_a\};L)=
\overline{Z({\bf R}_1,{\bf R}'_1;L)\cdots Z({\bf R}_n,{\bf R}'_n;L)}= 
\int_{{\bf R}_a(0)={\bf R}_a}^{{\bf R}_a(L)={\bf R}'_a} 
\prod_{a=1}^n[d{\bf R}_a]\exp(-\beta H_n),
\label{Zn}
\end{equation}
where
\begin{eqnarray}
H_n=\frac{1}{2}\int_0^L du \sum_{a} \left[ M\left(\frac{d {\bf{R}}_a (u)}{d 
u}\right)^2 +\mu {\bf{R}}^2_a (u) \right] \nonumber \\
-\frac{\beta}{2} \int^L_0 du 
\int^L_0 du' \sum_{ab} f\left( ({\bf{R}}_a (u)-{\bf{R}}_b (u'))^2 \right).
\label{Hn}
\end{eqnarray}
The averaged equilibrium properties of the polymer can now be written
in terms of the replicated partition sum ${Z_n}(\{{\bf{R}}_a\},\{{\bf{R}}_a\};L)$.
For instance, the mean squared displacement defined in Eq.~(\ref{rt2av}) can be 
written in as
\begin{equation}
\overline{{\langle} {\bf{R}}_T^2(L){\rangle}}=\lim_{n\rightarrow 0} \,\, 
\frac{\int d{\bf{R}}_1 
\cdots d{\bf{R}}_n {\bf{R}}^2_1 {Z_n}(\{{\bf{0}}\},\{{\bf{R}}_a\};L)}
{\int d{\bf{R}}_1 \cdots d{\bf{R}}_n 
{Z_n}(\{{\bf{0}}\},\{{\bf{R}}_a\};L)},
\label{wanderT}
\end{equation}
and similarly
\begin{equation}
\overline{{\langle} {\bf{R}}_F^2(L){\rangle}}=\lim_{n\rightarrow 0} \,\, 
\frac{\int \prod d{\bf{R}}_a
\prod d{\bf R}'_a \left({\bf R}_1 - {\bf R}'_1\right)^2 {Z_n}(\{{\bf{R}}_a\},
\{{\bf R}'_a\};L)}{\int \prod d{\bf R}_a \prod d{\bf R'}_a
{Z_n}(\{{\bf{R}}_a\},\{{\bf R}'_a\};L)}.
\label{wanderF}
\end{equation}
Thus, the averaged equilibrium properties of the polymer can be extracted from 
an $n$-body problem by taking the $n\rightarrow 0$ limit at the end. This limit
has to be taken with care, by solving the problem analytically for general $n$,
before taking the limit of $n\rightarrow0$. 

We now proceed to introduce our toy model that can be exactly solved using the 
replica method and which also lends itself to an accurate numerical solution.
This is the case when a Gaussian polymer chain is immersed in a random medium 
that has very long range spatial correlations.  In particular, we take the 
correlation function to be of the form
\begin{equation}
{\langle} V({\bf{R}})V({\bf{R}}'){\rangle}=f\left(({\bf{R}}-{\bf{R}}')^2 
\right)=g\left(1-({\bf{R}}-{\bf{R}}')^2/\xi^2\right),
\label{quadraticcor}
\end{equation}
where $\xi$ is chosen to be larger than the sample size, so that the 
correlation function is 
well defined (non-negative) over the entire sample.  Since this model 
can be solved exactly using the replica method, we can compute 
all the important physical properties of the polymer chain, and then 
compare the exact analytical results with an alternative numerical 
solution (at $d=1$). 
Also, this model of long range correlations is interesting in its own
right in that it may serve as a good approximation to any correlation
function $f$ that is smooth and slowly decaying. Most cases investigated 
so far in the literature
are concerned with disorder with short ranged correlations.    

There are many properties of the polymer chain that can be 
exactly computed.  In addition to $\overline{{\langle} {\bf{R}}_T^2(L){\rangle}}$
and $\overline{{\langle} {\bf{R}}_F^2(L){\rangle}}$ we will compute two 
other quantities. First, for a polymer loop of arc-length $L$,
we will compute the quantity \cite{gold,edwards}
\begin{equation}
C(l)=\frac{1}{d} \overline{\langle {\bf{R}}(l)-{\bf{R}}(0) \rangle
^2}=\frac{1}{d} \overline{\left(  \frac{
\int d{\bf{R}} d{\bf{R'}}
({\bf{R'-R}})^2 Z({\bf{R}},{\bf{R'}};L-l)Z({\bf{R'}},{\bf{R}};l)} 
{\int d{\bf{R}} Z({\bf{R}},{\bf{R}};L) }\right)},
\label{cl}
\end{equation} 
in the limit $L \gg l$.  This is a measure of the average fluctuations of
a chain segment of arc-length $l$. Since in this case the chain is not anchored,
this quantity is in some respect similar to $\overline{{\langle} 
{\bf{R}}_F^2(L){\rangle}}$. Yet another quantity of interest is 
\begin{equation}
\overline{{\langle} {{\bf{R}}^2_Q}(L){\rangle}}=\overline{\left(\frac{
\int d{\bf{R}} {\bf{R}}^2 Z({\bf{R}},{\bf{R}};L)}{\int d{\bf{R}} 
Z({\bf{R}},{\bf{R}};L)}\right)},
\end{equation} 
which has a more direct application to the related problem of a
quantum particle in a random potential \cite{gold2}.
The reason for this is that the partition sum of
a polymer chain can be mapped to the density matrix of a quantum particle.
The mapping \cite{gold2,feynman} is given by   
\begin{equation}
\beta \rightarrow 1/\hbar, \; \; L \rightarrow \beta \hbar.
\end{equation}
Then  $\rho(R,R';\beta)=Z(R,R';L=\beta \hbar,\beta=1/\hbar)$ is
the density matrix of a quantum particle at inverse temperature $\beta$.
Note that the variable $u$ is now interpreted as the Trotter (imaginary)
time, and $M$ as the mass of the quantum particle.  Under this
mapping $\overline{{\langle} {{\bf{R}}^2_Q}(L){\rangle}}$ can
be interpreted as the average mean squared displacement of a quantum
particle in a random plus harmonic potential. 

The paper is organized as follows: In Sec.~\ref{s2} we outline the exact
analytical solution for various quantities relevant to a polymer chain.
In the next section we present the details of the numerical approach 
to the problem.  In Sec.~\ref{s4} we compare the analytical and numerical 
results and comment on the physical implications of our results. 
Concluding remarks are offered in Sec.~\ref{s5}.

\section{The analytical solution}
\label{s2}

We start with the case when one end point is fixed.  The analytical 
calculation is based on an exact evaluation of the replicated
partition sum (\ref{Zn}).  For the correlation function $f$ that we are 
considering the replicated Hamiltonian is
\begin{equation}
H_n=\frac{1}{2}\int_0^L du \sum_{a} \left[ M\left(\frac{\partial {\bf{R}}_a
(u)}{\partial u}\right)^2 +\mu {\bf{R}}^2_a (u) \right]+ \beta\sigma\int^L_0 du 
\int^L_0 du' \sum_{ab} ({\bf{R}}_a (u)-{\bf{R}}_b (u'))^2 ,
\end{equation}
where $\sigma=g/2{\xi}^2$, and where we have dropped the constant part of 
the function $f$ since it only contributes an unimportant normalization 
factor.  After expanding the quadratic term and simplifying the double 
integral we get the replicated Hamiltonian
\begin{equation}
H_n=\frac{1}{2}\int_0^L du \sum_{a} \left[ M\left(\frac{\partial 
{\bf{R}}_a (u)}{\partial 
u}\right)^2 +(\mu+4n\beta\sigma L) {\bf{R}}^2_a (u) \right]- 
2\beta\sigma \left( \sum_{a} \int^L_0 du {\bf{R}}_a(u) \right)^2 . 
\end{equation} 
Now, using the Gaussian transformation 
\begin{equation}
e^{{\bf{Q}}^2/2}=\frac{1}{(2\pi)^{d/2}} \int_{-\infty}^{\infty} 
d{\mbox{\boldmath $\lambda$}} e^{(-{\mbox{\boldmath $\lambda$}}^2/2-{\bf{Q}} \cdot 
{\mbox{\boldmath $\lambda$}})}
\end{equation}
and letting
\begin{equation}
{\bf{Q}}=2 \beta\sqrt{\sigma} \left({\sum_{a} \int^L_0 du {\bf{R}}_a(u)} \right),
\end{equation}
we can write the replicated partition sum as 
\begin{equation}
Z_n(\{{\bf{R}}_a\},\{{\bf R}'_a\};L)=\frac{1}{{(2\pi)}^{d/2}}\int_{-\infty}^{\infty}
d{\mbox{\boldmath $\lambda$}} e^{-{{\mbox{\boldmath $\lambda$}}}^2 /2}
 \prod_{a=1}^{n}\int_{{\bf{R}}_a(0)={\bf R}_a}^{{\bf R}_a(L)=
{\bf R}'_a}[d{\bf{R}}_a] e^{-\beta H_a({\mbox{\boldmath $\lambda$}})},
\end{equation}
where
\begin{equation}
H_a({\mbox{\boldmath $\lambda$}})= \int_{0}^{L} du  \left[{  
 \frac{M}{2}\left(\frac{d {\bf{R}}_a (u)}{du}\right)^2 
+\frac{{\mu}'}{2} {\bf{R}}^2_a (u) 
+2\sqrt{\sigma} {\mbox{\boldmath $\lambda$}} \cdot {\bf{R}}_a(u) }\right], 
\end{equation}
and where ${\mu}'=\mu+4n\beta\sigma L$.  The path integrals can now be evaluated 
directly using well known results for quadratic Hamiltonians.  
The details of the calculation are given in the Appendix.  Once the 
partition sum is known we 
can directly evaluate the right hand side of Eq.~(\ref{wanderT}) by 
taking $n \rightarrow 0$ at the very end.  The result 
is 
\begin{equation}
\overline{{\langle} {\bf R}_T^2 
(L){\rangle}}=\frac{d}{\beta}\sqrt{\frac{1}{M\mu}}\tanh
\left(\sqrt{\frac{\mu}{M}}L\right)+
\frac{4\sigma d}{\mu
^2}\left(1-\frac{1}{\cosh\left(\sqrt{\frac{\mu}{M}}L\right)}\right)^2 ,
\label{meansq}
\end{equation}

We can also compute the averaged mean displacement square 
$\overline{{\langle} {\bf{R}}_T(L){\rangle}^2}$ (see Appendix).  We find that
\begin{equation}
\overline{{\langle} {\bf{R}}_T(L){\rangle}^2}=\frac{4\sigma d}{\mu
^2}\left(1-\frac{1}{\cosh\left(\sqrt{\frac{\mu}{M}}L\right)}\right)^2.
\label{meansq2}
\end{equation}  
This implies that the displacement from the average is 
\begin{equation}
\overline{{\langle} {\bf{R}}_T^2(L) {\rangle}-{\langle} 
{\bf{R}}_T(L){\rangle}^2}=\frac{d}{\beta}\sqrt{\frac{1}{M\mu}}\tanh
\left(\sqrt{\frac{\mu}{M}}L\right),
\end{equation}  
which is independent of disorder.  We will discuss the physical implications
of these results in a later section.

Considering now a chain that is free to move we calculate (see Appendix for details)
the quantity $\overline{{\langle} {\bf{R}}_F^2(L){\rangle}}$ using 
Eq.~(\ref{wanderF}). The result is
\begin{equation}
\overline{{\langle} {\bf R}_F^2 
(L){\rangle}}=\frac{2d}{\beta\sqrt{M\mu}}
\frac{\sinh\left(\sqrt{\frac{\mu}{M}}L\right)}
{\left(\cosh\left(\sqrt{\frac{\mu}{M}}L\right)+1\right)},
\label{rsqf}
\end{equation}
which is independent of disorder and in the limit of $\mu \rightarrow 0$ behaves
like $dL/\beta M$, i.e. like a free chain.

The quantity $\overline{{\langle} {\bf{R}}^2_Q (L){\rangle}}$  can also 
be computed exactly from the expression
\begin{equation}
\overline{{\langle} {\bf{R}}_Q^2(L){\rangle}}=\lim_{n\rightarrow 0} \,\, 
\frac{\int d{\bf{R}}_1 
\cdots d{\bf{R}}_n {\bf{R}}^2_1 {Z_n}(\{{\bf{R}}_a\},\{{\bf{R}}_a\};L)}
{\int d{\bf{R}}_1 \cdots d{\bf{R}}_n 
{Z_n}(\{{\bf{R}}_a\},\{{\bf{R}}_a\};L)}.
\label{wanderQ}
\end{equation}
Details are given in the Appendix. However, in this case it is also possible
to carry out the computation in an alternative way by taking advantage
of the 
periodic boundary conditions of the closed loop. This provides for a further 
check on the result and is also included for instructional purposes.
Using the Fourier space variables 
$${\bf{R}}_a(\omega)=(1/\sqrt{L})\int_0^L 
du {\bf{R}}_a(u)e^{-i\omega u},$$
we can write the propagator associated with $\beta H_n$ as
\begin{eqnarray}
\beta G_{ab} (\omega)&=&\frac{\beta}{d} \left\langle {\bf{R}}_a(\omega) \cdot
{\bf{R}}_b(-\omega)\right\rangle \nonumber \\ 
&=&\left\{  \left( M\omega^2 +\mu +4n\beta\sigma L \right) {\bf I} - 
4\beta\sigma L \delta_{\omega,0}
\right\} _{ab} ^{-1},
\end{eqnarray}
where $\omega$ is restricted to the discrete values
 \begin{equation}
\omega_m =\frac{2\pi}{L} m, \;\;  m=0,\pm 1,\pm 2,...\; \; .
\end{equation}
After inverting the $n \times n$ matrix and taking the $ n \rightarrow
0 $, we find 
\begin{equation}
\beta G_{ab} (\omega =0)=\frac{\delta_{ab}}{\mu}+\frac{4\beta\sigma L}{\mu^2},
\end{equation}
\begin{equation} 
\beta G_{ab} (\omega \neq 0)=\frac{\delta_{ab}}{M\omega^2+\mu}.
\end{equation}
Then, using the relation
\begin{equation}
\langle {\bf{R}}^2_a (L) \rangle =\frac{d}{L} \sum_{\omega}  G_{aa} (\omega),
\end{equation}
we find that
\begin{equation}
\overline{{\langle} {\bf{R}}^2_Q (L){\rangle}}=\frac{1}{n} \sum_{a=1}^{n}
\langle {\bf{R}}^2_a (L) \rangle
=\frac{4\sigma 
d}{\mu^2}+\frac{d}{2\beta\sqrt{M\mu}}\coth\left(\sqrt{\frac{\mu}{M}}
\frac{L}{2}\right),
\label{shift}
\end{equation}
which implies that the only effect of the disorder is to shift the
zero disorder result by a constant factor.  Next, we compute the
quantity $\overline{{\langle} {\bf{R}}_Q (L){\rangle}^2}$.  We
find that 
\begin{eqnarray}
\overline{{\langle} {\bf{R}}_Q (L){\rangle}^2} &=& \frac{1}{n(n-1)}
\sum_{a \neq b}^{n} \langle {\bf{R}}_a (L) \cdot  {\bf{R}}_b
(L)\rangle 
\nonumber \\
&=& \frac{d}{Ln(n-1)} \sum_{a \neq b}^{n} \sum_{\omega} G_{ab}
(\omega)=\frac{4\sigma d}{\mu^2},
\label{shift2}
\end{eqnarray}
which again implies that the deviation from the average 
$\overline{{\langle} {\bf{R}}^2_Q (L){\rangle}-{\langle} {\bf{R}}_Q
(L){\rangle}^2}$ is independent of disorder.

Finally, we compute the quantity $C(l)$, which was defined in Eq.~(\ref{cl}).
We use,
\begin{eqnarray}
C(l) &=& \frac{1}{nd} \sum_{a=1}^{n} \langle \left( {\bf{R}}_a (l)-{\bf{R}}_a
(0) \right) ^2 \rangle \nonumber \\
&=& \frac{2}{nL} \sum_{a=1}^{n} \sum_{\omega} G_{aa}(\omega)(1-e^{-i\omega l})
=\frac{2}{\beta L} \sum_{\omega \neq 0} \frac{1-e^{-i\omega
l}}{M\omega^2+\mu},
\end{eqnarray}
which for large $L$ yields the expression 
\begin{equation}
C(l)=\frac{1}{\beta \sqrt{M \mu}} \left( 1-{\exp}(-l\sqrt{\mu/M} \right).
\label{floop}
\end{equation}
So we find that $C(l)$ is independent of the disorder and is the same
as that of a free chain. 
 
\section{Numerical Procedure}
\label{s3}

In order to check the validity of the analytical solution we will have
to numerically compute the quenched average of certain physical properties
of the polymer.  This is a rather computationally intensive task
because of the difficulty of evaluating the partition sum, and also
because all quantities will then have to be averaged over many
realizations of the random 
potential.  In this paper we will only concentrate on the case
$d=1$.  Although this does not correspond to a physical polymer ($d=3$)
we will still be able to check  the validity of our analytical results for 
the special case $d=1$.   In the context of the quantum particle in
a random potential this case corresponds to a particle
in a one dimensional random potential.

We evaluate the path integral (\ref{Z}) numerically by mapping it to the
associated Schr\"odinger equation.  In dimension $d=1$ this 
mapping (see Ref.~{\cite{feynman}} Eqs.~(3.12)-(3.18)) is given by
\begin{equation}
Z(R,R';L)=\int_{R(0)=R'}^{R(L)=R} [dR(u)] {\exp}\left(-\beta H[R(u)]\right)= 
{\langle} R|{\exp}(-\beta L\hat{H})|R'{\rangle},
\label{map}
\end{equation}
where 
\begin{equation}
\hat{H}=-\frac{1}{2M\beta^2}\frac{\partial^2}{\partial
\hat{R}^2}+\frac{\mu}{2}\hat{R}^2 +V(\hat{R}).
\end{equation}
We compute the matrix element by expanding it in terms of the energy
eigenstates of $\hat{H}$
\begin{equation}
{\langle} R|{\exp}(-\beta L\hat{H})|R'{\rangle}=\sum_{n} {\exp}(-\beta L
E_n) \Phi_n(R)^* \Phi_n(R').
\label{eigen}
\end{equation}
In order to compute the eigenvalues and eigenvectors numerically we 
solve the Sch\"ordinger equation on a one dimensional lattice of 
$N$ sites \cite{press}.  The lattice Hamiltonian is then 
an $N\times N$ matrix with matrix elements given by 
\begin{equation}
H_{ij}=-\frac{1}{2M\beta^2\Delta^2} \left( \delta_{i,j+1}+\delta_{i+1,j} 
\right)  +
\left(\frac{\mu}{2}\Delta^2 (i-N/2)^2 + V(i) \right) \delta_{i,j}
\end{equation}
where the lattice spacing is $\Delta=S/N$, and where $S$ is the system
size.  Since we are interested in the continuum limit  $\Delta$ will be
kept small.  Note that the index $i$ corresponds to the 
position $R_i=\Delta i$.  The eigenvalues and eigenvectors can 
now be found directly by diagonalizing the matrix using a standard numerical
routine \cite{press}.  Once these are known we can construct the
partition sum at any value of $L$ using Eq.~(\ref{eigen}).  

The random potential $V(R)$ is generated by first 
generating a Gaussian correlated random potential $V_{\xi}(R)$
that satisfies
\begin{equation}
\langle V_{\xi}(R)V_{\xi}(R') \rangle \propto
{\exp}\left({- (R-R')^2/\xi^2}\right).
\label{Gaussiancor}
\end{equation} 
Since we are making a lattice approximation we need a sequence 
of $N$ numbers $\{ V_{\xi}(i) \}_{i=1,..,N}$ that 
obey $\langle V_{\xi}(i)V_{\xi}(i+l) \rangle \propto 
G(l)$, where in this case  
$G(l)= {\exp}\left(- \Delta^2 l^2/\xi^2 \right)$.  
These numbers will then be placed on the $N$ lattice sites in the 
given order.  To generate such numbers we use a method described in reference 
\cite{stanley}.  The procedure is to first generate a sequence
of $N$ uncorrelated random numbers $\{U(i)\}$ with a Gaussian distribution.
These numbers are then fast Fourier transformed, using a standard
numerical routine \cite{press}, to yield the sequence $\{ \widetilde{U}(i) \}$.
Next, we calculate the $N$ numbers defined 
by $\widetilde{W}(i)=\sqrt{\widetilde{G}(i)}\widetilde{U}(i)$,
where $\widetilde{G}(i)$ is defined as the  Fourier transform of the 
correlation function $G(i)$. Finally, taking the inverse 
Fourier transform of the sequence $\{\widetilde{W}(i)\}$, 
yields $ \{W(i)\} $, the 
sequence with the desired correlation function $G(i)$.
Now, in order to generate quadratic correlations we
choose $\xi$ such that the Gaussian correlation
function is well approximated by its leading quadratic term over the
range of the system size. The approximate condition for this to hold is 
that $\xi/S \gtrsim 1/\sqrt{2} $.   In this way we generate a well defined 
set of random numbers which obey approximately the correlation function 
given in Eq.~(\ref{quadraticcor}).  

\begin{figure}
\centerline{\epsfysize 8.5cm \epsfbox{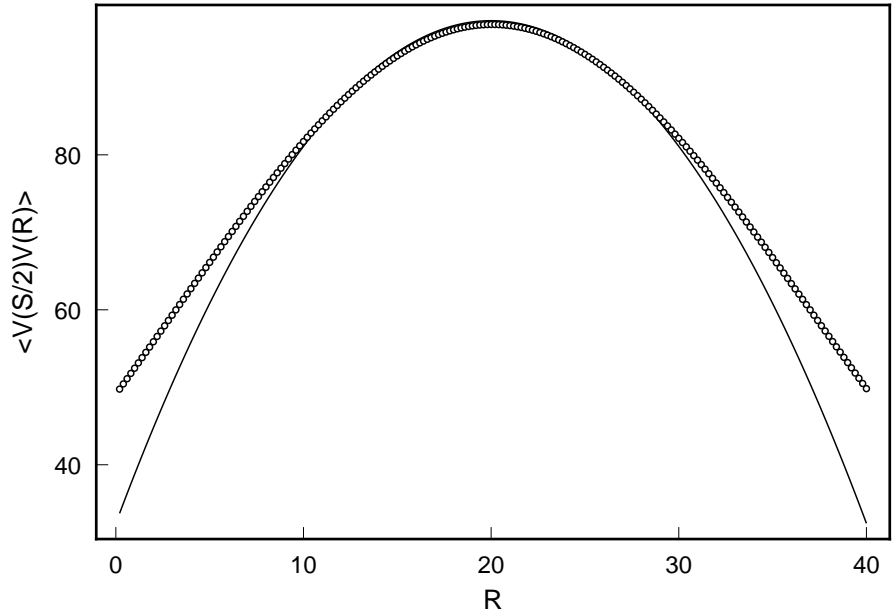}} \vspace{3mm}
\caption{Plot of  ${\langle} V(S/2)V(R){\rangle}$ versus 
$R$. The circles are generated by averaging over $10,000$ samples.
The solid line is a plot of the quadratic approximation to the 
correlation function given by Eq.~(\ref{Gaussiancor}).  The parameters 
are $\xi=10 \sqrt{6}$, $S=40$, N=200.}  
\label{cor}
\end{figure}

In Fig.~{\ref{cor}} we plot a 
correlation function that is generated by the above method.  On the
same graph we plot the corresponding quadratic approximation.
Notice that in this case when $\left|R-S/2 \right| \sim 10$ the quadratic 
approximation begins to deviate from the generated correlation
function.  This discrepancy turns out to be unimportant as long 
the quantities that are numerically computed (such as end to end distance) do 
not exceed this range of validity. 
In order to reduce errors due to the finite size of the lattice we found it
useful to take our sample of random numbers from a set which was 
about five times $N$.  In all cases we tested the reliability of
the samples by directly computing the correlation function and
comparing to the analytical expression for the correlation function given by 
Eq.~(\ref{Gaussiancor}).

\section{Results and Discussion}
\label{s4}

We discretized the Schr\"odinger equation on a lattice 
of size $N=200$. Once the eigenvalues and eigenvectors are 
known then we can approximate, for instance the mean squared 
displacement, for each random sample.  We then average over 
the samples to get an approximation to the quenched average.
For simplicity we set $M=1/2$ and $\beta=1$ for all cases. 

\begin{figure}
\centerline{\epsfysize 8.5cm \epsfbox{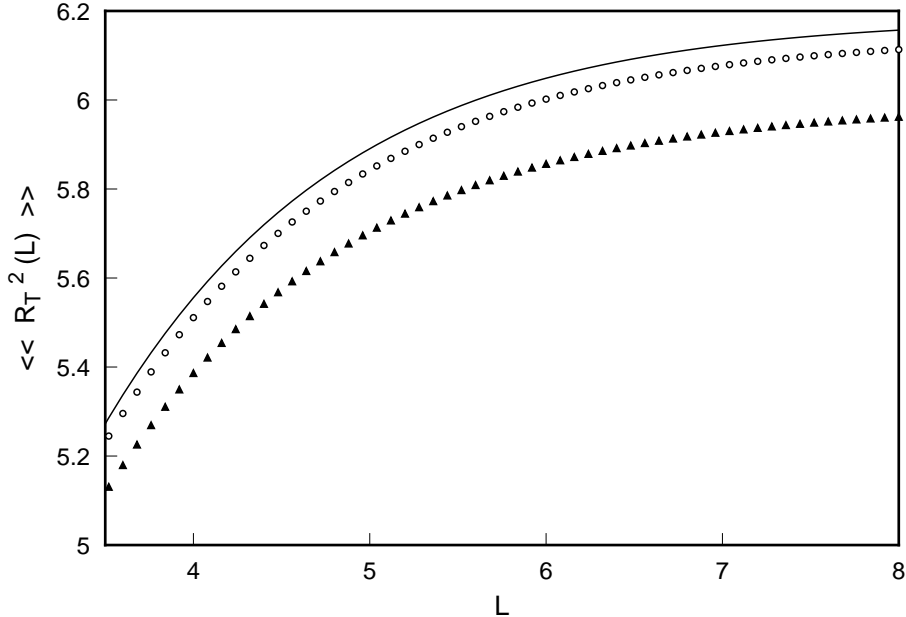}} \vspace{3mm}
\caption{Plot of $\overline{{\langle} R_T^2 (L){\rangle}}$ as a
function of $L$.  The parameters are $\mu=0.3$, $\sigma=0.0811$,
$\Delta=0.2$, $\xi =10 \sqrt{6}$. The solid line corresponds to the
analytical solution given in Eq.~(\ref{meansq}). The triangles are
generated by averaging over $1000$ samples and the circles represent
averaging over $10,000$ samples.}
\label{fix1}
\end{figure}

\begin{figure}
\centerline{\epsfysize 8.5cm \epsfbox{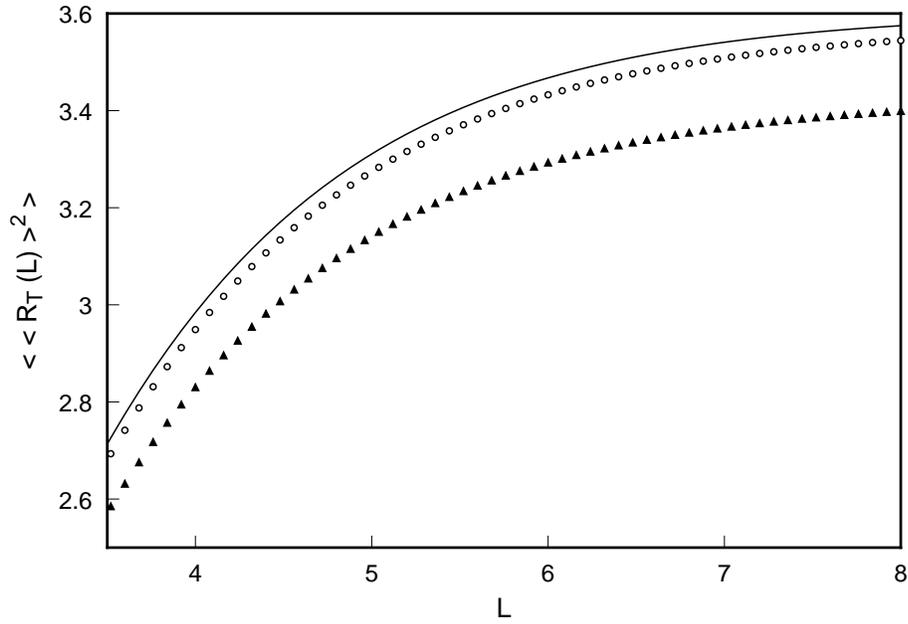}} \vspace{3mm}
\caption{Plot of $\overline{{\langle} R_T (L){\rangle}^2}$ as a function
of $L$.  The solid line is a graph of the analytical solution in
Eq.~(\ref{meansq2}). The triangles are
generated by averaging over $1000$ samples and the circles represent
averaging over $10,000$ samples.}
\label{fix2}
\end{figure}

\begin{figure}
\centerline{\epsfysize 8.5cm \epsfbox{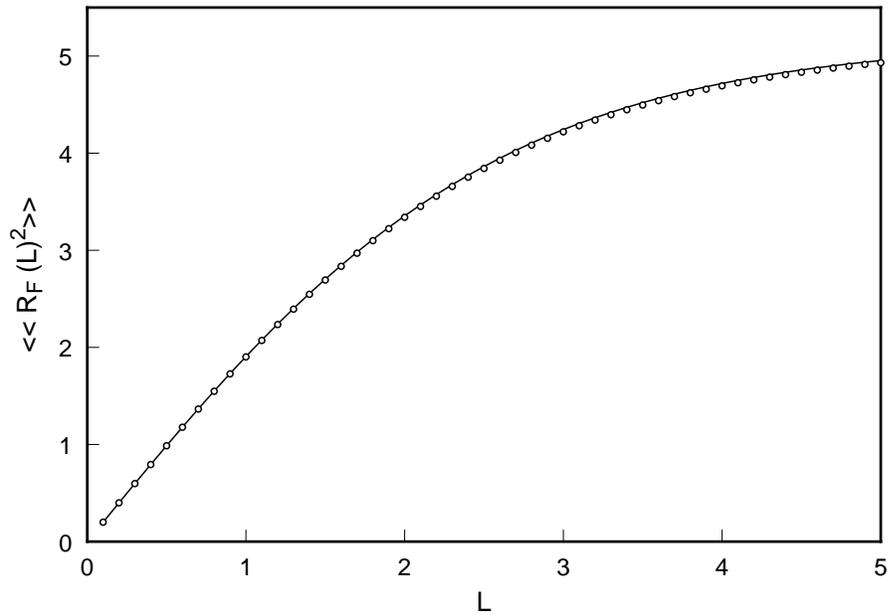}} \vspace{3mm}
\caption{Plot of $\overline{{\langle} R_F^2 (L){\rangle}}$ as a
function of $L$.  The solid line corresponds to the
analytical solution given in Eq.~(\ref{rsqf}). The circles are
generated by averaging over $200$ samples. All the parameters are the
same as in Fig~\ref{fix1}.}
\label{free}
\end{figure}

In Fig.~{\ref{fix1}} we graph the mean squared displacement with 
one endpoint fixed as a function of $L$.  We do this for two different
number of samples in order to check convergence towards the 
corresponding analytical 
solution. Note that in the labels of the plots the average over the 
disorder is denoted by a second set of brackets rather than an overbar.  
In Fig.~{\ref{fix2}} we graph $\overline{{\langle} R_T (L){\rangle}^2}$ as a
function of $L$. We use the same parameters as in Fig.~{\ref{fix1}}.
It is clear from the graphs that the numerical results are consistent
with the exact curve.  As the number of the samples is increased the 
numerical curves get closer to the analytical solution.  
In Fig.~{\ref{free}} we plot $\overline{{\langle} R_F^2 (L){\rangle}}$
vs. $L$.  This quantity is computed numerically using the expression
given in Eq.~(\ref{rf2av}).  We found that the numerical results were
extremely close to the analytical prediction after averaging over only
$200$ samples. 


\begin{figure}
\centerline{\epsfysize 8.5cm \epsfbox{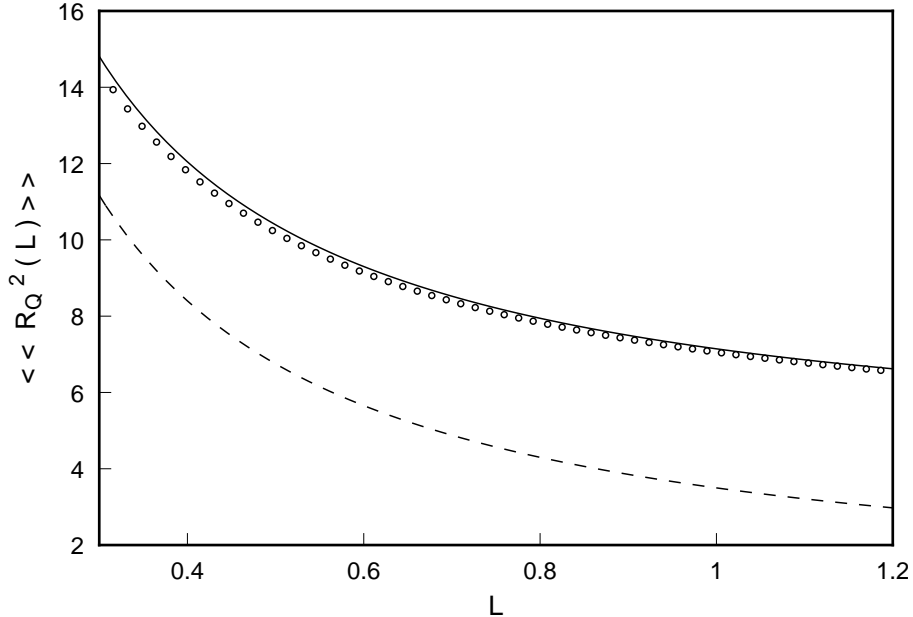}} \vspace{3mm}
\caption{Plot of $\overline{{\langle} R^2_Q (L){\rangle}}$ vs. $L$.
The solid line is the analytical solution given in Eq.~(\ref{shift}).  The
circles are generated by averaging over $8000$ samples.  The dashed
line is the analytical solution for zero disorder ($\sigma=0$).}  
\label{period1}
\end{figure}

\begin{figure}
\centerline{\epsfysize 8.5cm \epsfbox{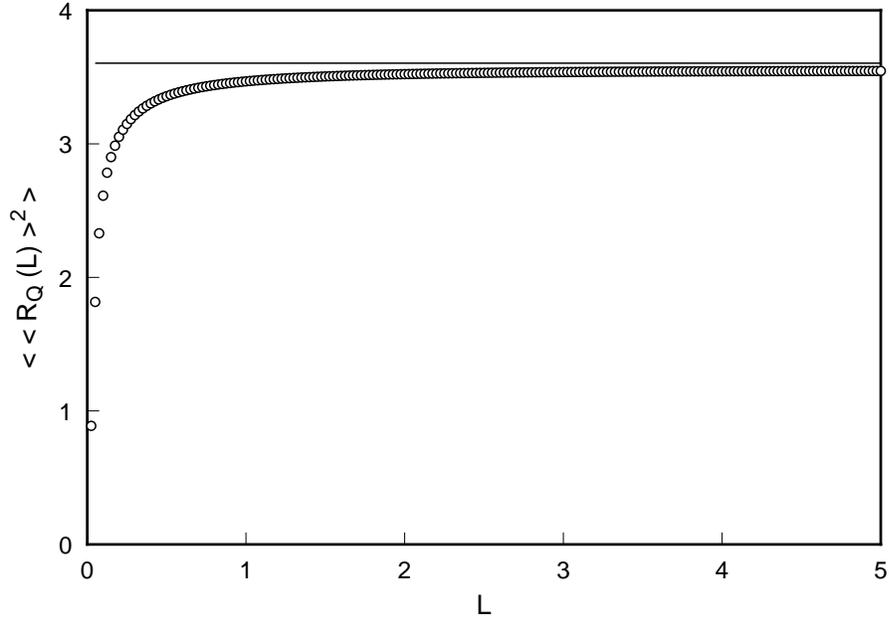}} \vspace{3mm}
\caption{Plot of $\overline{{\langle} R_Q (L){\rangle}^2}$ vs. $L$.
The solid line is the analytical solution given in Eq.~(\ref{shift2}).  The
circles are generated by averaging over $8000$ samples.}  
\label{shift3}
\end{figure}

We now turn our attention to the quantity 
$\overline{{\langle} R^2_Q (L){\rangle}}$, which was
discussed in the introduction.  In Fig.~{\ref{period1}}. we 
graph $\overline{{\langle} R^2_Q (L){\rangle}}$ vs. $L$.  
In order to visualize the predicted shift
in Eq.~(\ref{shift}) we include the exact solution of the 
zero disorder case. We use the same parameters as the previous 
figures.
Again, we find close agreement between the computational results
and the analytical solution.  The shift due to the disorder is clearly
evident and is very close to the predicted value.  In Fig.~{\ref{shift3}}
we plot  $\overline{{\langle} R_Q (L){\rangle}^2}$ vs. $L$ and compare
with a plot of the analytical solution in Eq.~(\ref{shift2}). 
For small $L$ there appears to be a discrepancy between the data
and the analytical solution, whereas for larger $L$ the two curves are
very close. This is due to the fact that the random potential is generated on 
a grid with grid size of 0.2.
Thus for L shorter than 0.2 the particle can not see the random potential and 
$\overline{{\langle} R_Q (L){\rangle}^2}$ vs. $L$ should average to zero. Indeed
the significant deviation occurs on this length scale. 

\begin{figure}
\centerline{\epsfysize 8.5cm \epsfbox{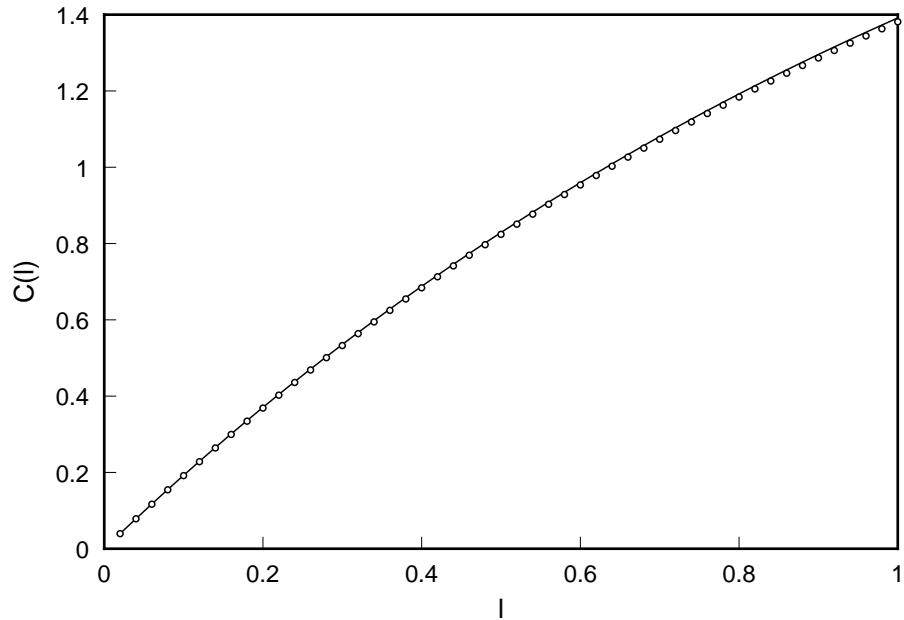}} \vspace{3mm}
\caption{Plot of $C(l)$ vs. $l$, with $L=10$.  The solid line is 
the analytical solution given by Eq.~(\ref{floop}).  The circles are 
generated by averaging over $250$ samples.  All parameters
are the same as the previous graphs.}  
\label{loop1}
\end{figure}

Finally, we turn our attention to the quantity $C(l)$. It was evaluated 
numerically from the expression given in Eq. (\ref{cl}). In
Fig.~{\ref{loop1}} we plot $C(l)$ vs. $l$ with $L$ large and
fixed.  

It is clear from the graphs that the numerical solutions are 
consistent with the exact analytical solution.
As expected, as the number of samples is increased the numerical results
get closer and closer to the exact curve.  Based on these results 
we can safely conclude that the replica calculation is indeed correct and  
does describe the averaged properties of the polymer.  

\begin{figure}
\centerline{\epsfysize 8.5cm \epsfbox{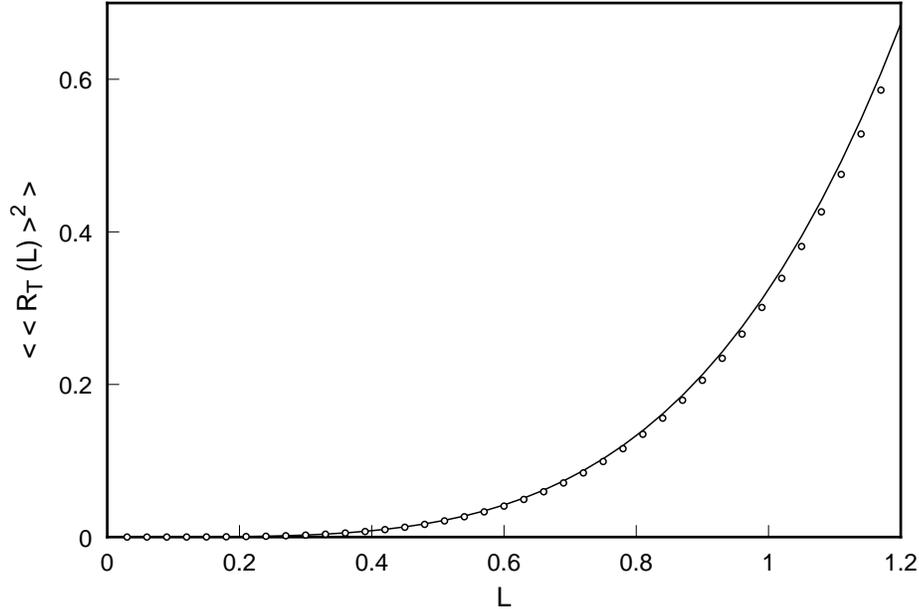}} \vspace{3mm}
\caption{Plot of $\overline{{\langle} R_T (L){\rangle}^2}$ vs. $L$.
The solid line is the analytical solution given in Eq.~(\ref{scaling}).
The circles are generated by averaging over $1000$ samples.
We take $\mu=0.001$, and all other parameters are the same as in
Fig.~{\ref{fix1}}.}  
\label{small1}
\end{figure}

\begin{figure}
\centerline{\epsfysize 8.5cm \epsfbox{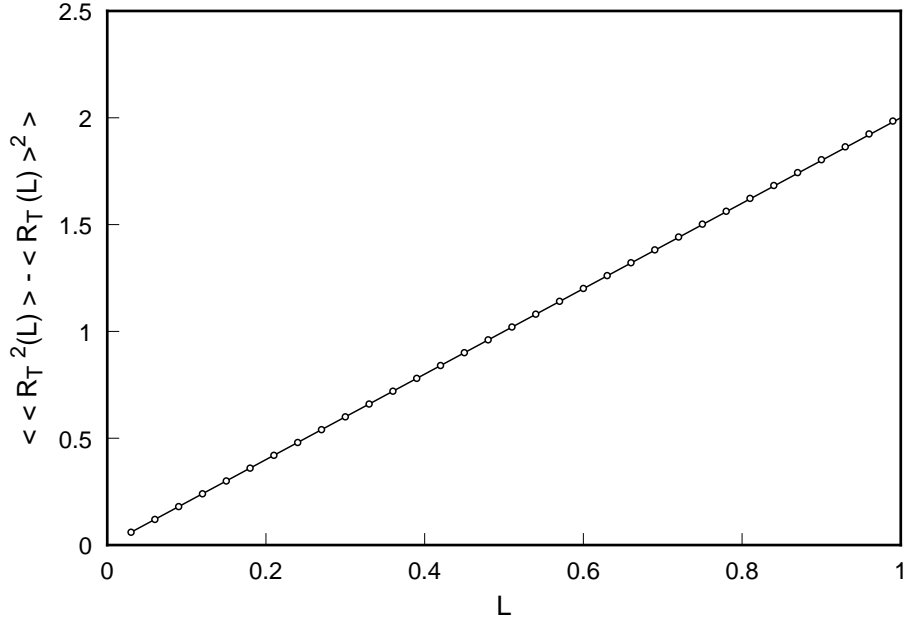}} \vspace{3mm}
\caption{Plot of $\overline{{\langle} R_T^2 (L) {\rangle}-{\langle}
R_T (L){\rangle}^2}$ vs. $L$.
The solid line is the analytical solution given by subtracting
Eq.~(\ref{scaling}) from Eq.~(\ref{smallmu}).
The circles generated by averaging over $1000$ samples.
We take $\mu=0.001$, and all other parameters are the same as in
Fig.~{\ref{fix1}}. }  
\label{samllmu3}
\end{figure}

It is interesting to study the infinite volume limit $\mu \rightarrow
0$.  Here, the polymer does not see the confining harmonic potential
and its properties are determined only by the random potential.
Taking the  $\mu \rightarrow 0$ limit the exact expressions simplify to
 \begin{equation}
\overline{{\langle} {\bf{R}}_T^2 (L){\rangle}}=\frac{d}{\beta
M}L+\frac{d\sigma}{M^2}L^4 ,
\label{smallmu}
\end{equation}
\begin{equation}
\overline{{\langle} {\bf{R}}_T^2 (L){\rangle}^2}=\frac{d\sigma}{M^2}L^4.
\label{scaling}
\end{equation} 
Notice that in the no disorder case $(\sigma=0)$ the later quantity
is zero, but once the disorder is turned on it scales like $L^4$ with 
a coefficient that is independent of temperature.
Also, Eq.~(\ref{smallmu}) indicates that for small $L$ the 
polymer wanders diffusively but for larger $L$ it wanders much faster 
than diffusion. In Fig.~{\ref{small1}} we plot $\overline{{\langle}
R_T (L){\rangle}^2}$  vs. $L$ when $\mu$ is chosen to
be very small.  On the same graph we plot Eq.~(\ref{scaling}). 
In Fig.~{\ref{samllmu3}} we plot  
$\overline{{\langle} R_T^2 (L) {\rangle}-{\langle}
R_T (L){\rangle}^2}$  vs. $L$.  On the same graph we include the
analytical prediction.
It is clear that the numerical results agree well with the 
analytical predictions.  Also, from Eq.~(\ref{shift}) we see that the 
quantity $\overline{{\langle} {\bf{R}}^2_Q (L) {\rangle}}$
diverges as $\mu \rightarrow 0$.  This implies that the boundary
conditions on the chain are crucial in determining which quantities
are well defined in the infinite volume limit. 

\begin{figure}
\centerline{\epsfysize 8.5cm \epsfbox{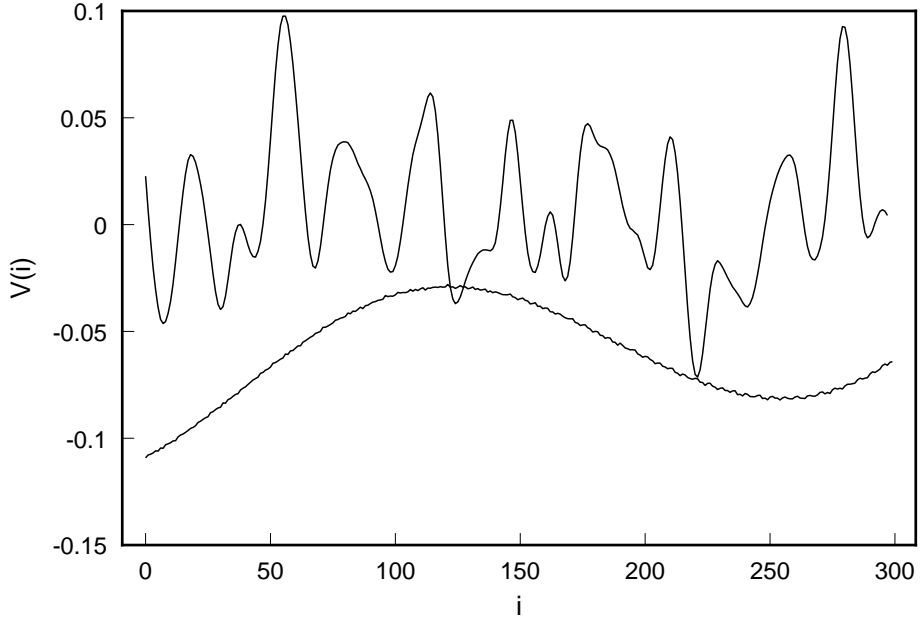}} \vspace{3mm}
\caption{The lower curve represents a typical random potential with long 
range quadratic correlations ($\xi=10\sqrt{6}$).  For comparison 
the upper curve represents a random potential with a shorter correlation range 
($\xi=1$).}
\label{rpot}
\end{figure}

The physical consequences of our results are surprising
and may seem counter intuitive at first glance.  The very long range 
correlations of the random
potential lead to a very fast wandering of the free end of a tethered 
chain. However, the deviation from the average position 
$\overline{{\langle} {\bf{R}}_T^2 (L) {\rangle}-{\langle}
{\bf{R}}_T (L){\rangle}^2}$  does not depend on the
random potential.  Also, if both ends are
free to move then the end to end distance 
$\overline{{\langle} {\bf{R}}_F^2 (L) {\rangle}}$ behaves as if there
is no random potential.  This behavior can
make more sense if we study the nature of the random 
potential samples that satisfy the quadratic correlation.
We find that the typical random potential (see Fig.~\ref{rpot})
is smooth and slowly varying on short scales 
but contains peaks and valleys on scales close to the system
size. 
So as $L$ is increased the polymer has a greater tendency 
to be found in the deepest
potential well in the sample, which, for large sample sizes,
is on average located very far from the fixed end of the polymer.
So for the case when one end is fixed we expect the end to end
distance to grow very fast with $L$, 
as the bulk (center of mass) of the polymer moves far away from 
the fixed end.  This behavior is the same as in the case of short 
range correlations
\cite{cates}, where the polymer will typically form a tadpole conformation
with the tail tethered to the origin and the head far away
in some region of low potential.  On the other hand, when both ends
are free the entire polymer will simply curl up in the region of
low potential and the end to end distance should depend only on
the local behavior of the random potential.
Now, since the potential samples are smooth and
slowly varying on short scales we do not expect the disorder to have
much of an effect on the local behavior of the polymer.
What is very interesting though is  
the fact that the chain that is free to move behaves as if the random 
potential has no effect at all. Similarly, the fluctuations around the average
position in the case when one end is fixed turn out to be totally independent of
disorder. This is expected to be a special feature of the quadratically 
correlated random potential, and is not likely to hold in the general
case of long range correlations. See discussion below about the crossover
to shorter ranged correlations of the disorder.

It is useful to compare our results with those for 
directed polymers. Here, the arc-length $u$ corresponds to the distance
along the directed axis which is a fixed direction in real space, 
and the vector ${\bf{R}}(u)$ is the position of
the directed line in the transverse hyper-plane. In that case the random 
potential is usually taken to depend on $u$ and satisfy 
\begin{equation}
\langle V({\bf{R}},u)V({\bf{R}}',u') \rangle = \delta (u-u')
f\left(({\bf{R}}-{\bf{R}}')^2 \right),
\label{dp}
\end{equation}
i. e. the random potential is taken to be uncorrelated along
the directed axis, unlike the situation described in Eqs. (\ref{Ham},\ref{VV})
were the random potential is independent of $u$. 
Parisi has shown  \cite{parisi} that if $f$ is quadratic, then the
mean squared displacement of one end of the directed polymer 
satisfies  $ \overline{\langle {\bf{R}}_T^2 (L) \rangle}  \propto L^3 $ .
This is to be compared with the $L^4$ dependence that we have found for
the case when the random potential is independent of $u$.

To explain the different scaling we employ a Flory-type argument similar
to the argument used in \cite{MP} for directed polymers. Allowing for a 
rescaling of the arc-length variable $u$ by a scale $\ell$, i.e. $u\ \rightarrow
\ell u$ and the position variables ${\bf R}(u)\ \rightarrow\ \ell^\zeta {\bf R}(\ell u)$
we see that the random potential which satisfies Eq.~(\ref{VV}) with a correlation
function behaving in general like
\begin{eqnarray}
f\left(({\bf{R}}-{\bf{R}}')^2\right)\ \sim\ const.-\frac{2\sigma}{1-\alpha}
\left({\bf{R}}-{\bf{R}}'\right)^{2(1-\alpha)},
\end{eqnarray}
scales like $\ell^\lambda$ with 
\begin{eqnarray}
2 \lambda=\zeta \ 2(1-\alpha)
\label{lambda}
\end{eqnarray}
The difference with directed polymers is that in that case one has to subtract 
a 1 from the right hand side of Eq.~(\ref{lambda}) because of the delta function
in Eq~(\ref{dp}).
Now in a Flory argument one assumes that the two terms in the hamiltonian given
in Eq.~(\ref{Ham}) scale the same way (here we consider only the case of $\mu=0$,
since $\mu \neq 0$ breaks scale invariance).
Since the 'kinetic' energy term scales like $\ell^{2\zeta-2}$, and this should
be equal to $\ell^\lambda$, we see that
\begin{eqnarray}
\zeta =\frac{2}{1+\alpha}.
\label{zeta}
\end{eqnarray}
Thus $\overline{{\langle} {\bf{R}}_T^2 (L){\rangle}}\sim L^{4/(1+\alpha)}$
as opposed to $L^{3/(1+\alpha)}$ for directed polymers.
In the quadratic case $\alpha=0$, and we get the $L^4$ behavior we were 
looking for. Notice that we derived here a prediction for the behavior
of $\overline{{\langle} {\bf{R}}_T^2 (L){\rangle}}$ for the case of long
ranged correlations of the disorder which are not quadratic but are characterized
by a power law determined by the value of $\alpha$. Thus the power of $L$ 
will decrease for shorter ranged correlations than quadratic. It is interesting to
assess the accuracy of the Flory argument in practice. In the function
$\overline{{\langle} {\bf{R}}_T^2 (L){\rangle}-\langle{\bf{R}}_T (L){\rangle}^2}$ or in 
$\overline{{\langle} {\bf{R}}_F^2 (L){\rangle}}$ the leading power of $L$ cancells
out and one is left with a subleading $L^1$ behavior (for quadratic correlations).

Within this toy model it is interesting to compare the differences
between the annealed and the quenched averages.  The annealed average
applies when the obstacles in the medium are randomly placed and 
mobile.  In this case the replica trick is not necessary and
the random potential can be averaged directly. Alternatively one can use the 
results obtained with $Z_n$ but instead of taking $n$ to 0, 
substituting $n=1$. 
We  easily find that
\begin{equation}
\overline{{\langle} {\bf{R}}_T^2 (L){\rangle}}=\frac{\int d{\bf{R}} {\bf{R}}^2 
\overline{Z({\bf{0}},{\bf{R}},L)}}{\int d{\bf{R}}
\overline{Z({\bf{0}},{\bf{R}},L)}} \sim \frac{1}{\sqrt{\sigma L}}\ ,
\end{equation}
when $L$ is large, and where we have taken the $\mu \rightarrow 0$
limit.  So in an annealed medium with long range quadratic
correlations a very long polymer chain will collapse around the tethered
end.  Similarly, we find that $\overline{{\langle} {\bf{R}}_F^2 (L){\rangle}}
\sim 1/\sqrt{\sigma L}$, and so if both ends are free then the polymer 
will collapse in the same way.   This behavior is in stark contrast to
the quenched case where the effects of the random medium is quite
different. 

For the case of short ranged correlations and a chain that is free 
to move one usually argues \cite{cates} that the annealed and quenched averages 
coincide in the infinite volume limit. This is due to the fact that the system 
can be divided into subregions, much larger than the chain, each containing 
a different realization of the potential. The moving chain can sample 
all of these and find a realization very similar to the one it induces 
around itself in the annealed case. But this
argument does not apply to the case of long ranged correlations
of the random potential, with the 
correlation length larger than the system size, since such a division to
subregions will not yield independent realizations.

It will be interesting to investigate how the physical properties
that we have found change as we move towards the regime of short
range correlations.  In our model we can control the correlation
length by varying the parameter $\xi$ in Eq.~(\ref{Gaussiancor})
i.e. in the Gaussian form. For small $\xi$ the correlation is 
certainly not quadratic, and it approaches a $\delta$-function 
in the limit $\xi \rightarrow 0$.  
For arbitrary $\xi$ we expect that the 
average mean displacement squared in $d=1$ will scale as  
$\overline{ \langle R_T (L) \rangle ^2} \propto  L^{\gamma (\xi)}$.  
Numerically, we can estimate the exponent $\gamma (\xi )$ by
performing a linear fit to the plot of 
$\log{\overline{\langle R_T (L) \rangle ^2}}$ vs. $\log{L}$
and measuring the slope.
For short range correlations we found that the numerical
method described in Sec.~\ref{s3} was unreliable.  The reason for this
is that the sum over energy eigenfunctions in Eq.~(\ref{eigen}) is unstable 
since a typical overlap $\Phi_n (R)^*\Phi_n (R')$ (for
short range correlations) is a number
on the order of $10^{-15}$.  However, we were able 
evaluate Eq.~(\ref{eigen}) 
accurately for all $\xi$ by solving
the Schr\"odinger equation on a lattice using  a fourth
order Runge-Kutta algorithm with a very small time step ($t \sim 10^{-4}$).  
We found that for large $L$ the quantity 
$\overline{ \langle R_T (L) \rangle ^2}$  saturates at a constant 
value due to the finite size of the system,  and so we do not 
expect a power law  scaling for large $L$.  However, 
for $L$ sufficiently small (before the onset of saturation) the 
mean displacement squared does obey a power law and a linear fit 
on a log-log plot was excellent for all $\xi$.    

\begin{figure}
\centerline{\epsfysize 8.5cm \epsfbox{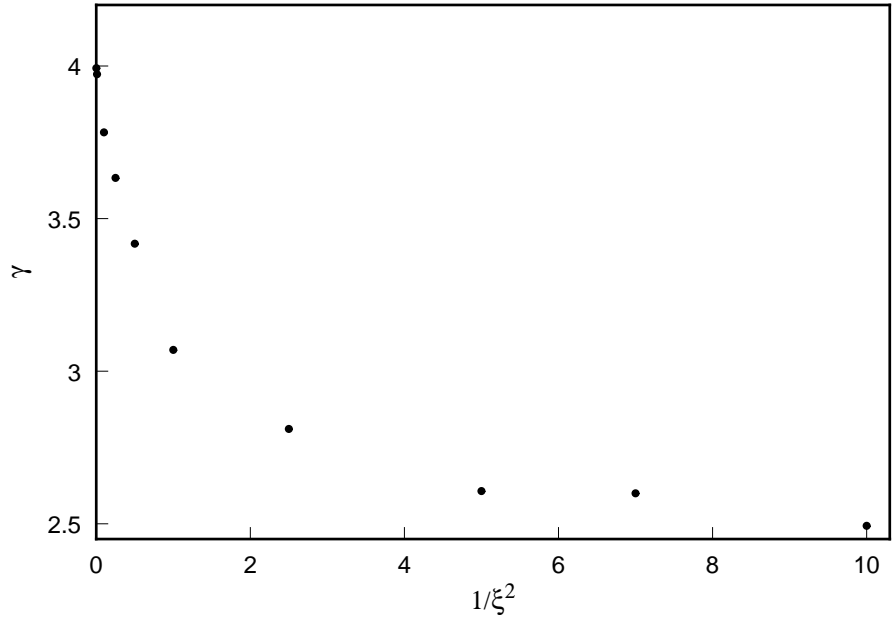}} \vspace{3mm}
\caption{Plot of $\gamma$ vs. $1/{\xi}^2$.}
\label{xi1}
\end{figure}

In Fig.~{\ref{xi1}} we plot 
$\gamma$ vs. $1/{\xi}^2$ for a range of
$\xi$.  For each point we averaged over $8000$ samples on a lattice
of size $N=300$, and in all
cases the strength of the random potential is taken to be 
large ($g \gg 1$). 
We can see from the plot that $\gamma$ falls from $4$, in the
case of very long range correlations, to about $2.5$ for very
short range correlations.  The case of delta correlated random
potentials has been studied by Nattermann
\cite{nattermann} using Flory arguments.  Nattermann finds that 
for strong disorder ($g \gg 1$)
the mean squared displacement behaves like
$\overline{ \langle {R}_T (L)^2 \rangle} \propto L^2
\sqrt{g} \left(\ln (L)\right)^{-3/2}$ in $d=1$. 
It is clear from Nattermann's arguments that 
$\overline{ \langle {R}_T (L)^2 \rangle} \sim 
\overline{ {\langle R_T (L) \rangle}^2}$, and so it is safe to compare
our numerical results with his analytical expression.
So while we find a scaling that is slightly faster than balistic
($\sim L^{2.5}$), Nattermann finds a weakly subballistic 
behavior ($\sim L^2 {\ln (L)}^{(-3/2)}$).
Nevertheless, it is comforting to see
that both results are fairly close to a balistic scaling 
($\sim L^2$).  

We now turn our attention to the chain that is free to move. 
Here, we find that for short range correlations ($\xi  \lesssim  \sqrt{5}$)
the end-to-end distance rises linearly for small $L$
and saturates at a constant value for large $L$.  This saturation
is not due to the finite size of the system since it occurs at a value
of $L$ far less than the length at which a free chain would saturate.
Typically, for $L  \gtrsim 1$ we find that 
$\overline{ \langle R_F^2 (L) \rangle} \propto  L^{0}$ as compared
to $\overline{ \langle R_F^2 (L) \rangle} \propto L $ when the
correlations are long range and quadratic.
In order to quantify this crossover between the long and short range
behavior we assume that the scaling relation $\overline{ \langle R_F^2 (L)
\rangle} \propto  L^{\delta (\xi)}$ holds for $L  \gtrsim 1$.  
Again we can estimate the exponent $\delta (\xi )$ by
measuring the slope of the line in a linear fit of 
$\log{\overline{\langle R_F^2 (L) \rangle }}$ vs. $\log{L}$.

\begin{figure}
\centerline{\epsfysize 8.5cm \epsfbox{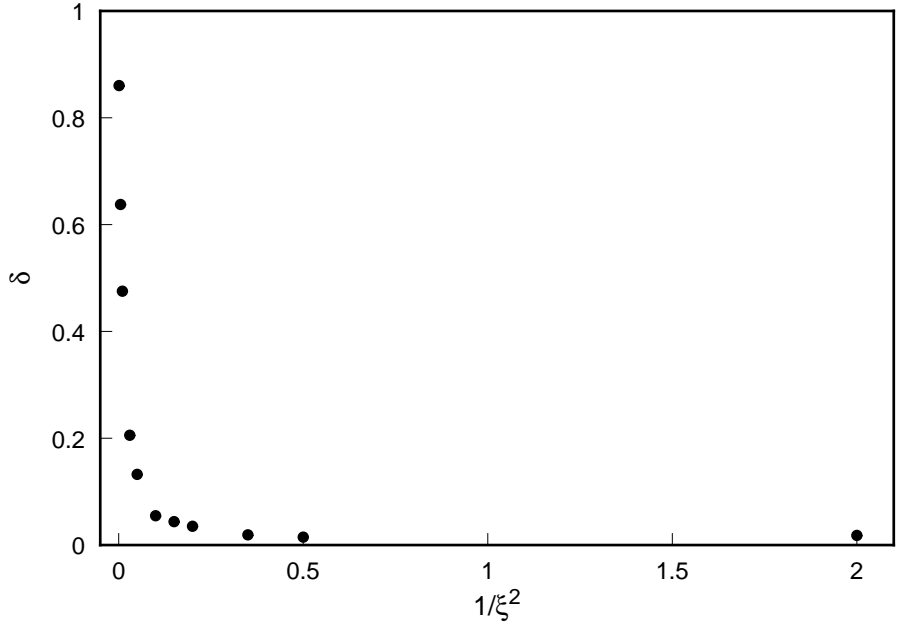}} \vspace{3mm}
\caption{Plot of $\delta$ vs. $1/{\xi}^2$.}
\label{xi2}
\end{figure}

In Fig.~{\ref{xi2}} we plot 
$\delta$ vs. $1/{\xi}^2$ for a range of
$\xi$.  
We computed the end-to-end distance for $L$ in the range $5<L<10$. 
For each point we averaged over $1000$ samples on a lattice
of size $N=200$, and in all cases the strength of the random potential
is taken to be large ($g \gg 1$).
We can see from the graph that as the correlation range is decreased 
$\delta$ falls rapidly from about $1$ to a value close to zero.
This implies that the behavior of $\overline{ \langle R_F^2 (L) \rangle}$
is very strongly dependent on the correlation range.
These results are consistent with the Flory arguments in
Ref.~\cite{cates,nattermann} where it is predicted that a long polymer
in a delta correlated random potential will have fixed size i.e. $R^2 \sim
L^0 $ in a sample of finite volume, and with the variational results of 
Ref.~\cite{gold} (see also \cite{edwards,mut}).

\section{Concluding Remarks}
\label{s5}

In this paper we presented a model of a polymer chain in a quenched random
media which was exactly solvable using the replica method.  
The analytical results were subsequently found to be in close agreement with a
numerical solution in $d=1$.  Based on these results we can safely
conclude that the replica method is accurate in describing the averaged 
properties of the polymer. The physical picture that emerged was interesting
and somewhat surprising.  We found that a quadratically correlated disorder 
has a major effect on the size of a polymer with one end fixed, 
but has no effect on the size of a chain that is free to move and 
find an optimal position. We also found that the quenched 
and annealed cases are rather different: In the annealed case 
a long chain collapses to a point.

Overall, we have learned that chain properties depend strongly on 
the correlation range of the random media.
However, there are still some open problems.  
For instance, it would be very useful to have an analytical derivation
of the results depicted in Figs.~{\ref{xi1},\ref{xi2}}.  Also, it may
be fruitful to investigate various non-equilibrium properties of
polymer chains in long range correlated random media, such as transport properties
and chain dynamics. We hope that our results for the 
simple case of  quadratic correlations will be a useful starting point 
for a more detailed analysis of these problems.
\begin{acknowledgments}
 Y.Y.G. acknowledges support from the US Department of Energy (DOE), grant No.
 DE-G02-98ER45686.
\end{acknowledgments}

\newpage
\appendix
\section*{}

\newcommand{\BR}{{\bf{R}}}
\newcommand{\BL}{{\mbox{\boldmath $\lambda$}}}
\newcommand{\BA}{{\bf{a}}}

Here we show some of the intermediate steps that lead to Eq.
(\ref{meansq}).  We first write the replicated partition sum as 
\begin{equation}
Z_n (\{\BR_a\},\{\BR'_a\};L)=\frac{1}{(2\pi)^{d/2}} \int d\BL
e^{-\BL^2/2} Z(\{\BR_a\},\{\BR'_a\};L,\BL),
\label{a1}
\end{equation} 
where 
\begin{equation}
Z(\{\BR_a\},\{\BR'_a\};L,\BL)=\prod_{a=1}^n \int_{\BR_a(0)=\BR_a}^{\BR_a(L)=\BR'_a}
[d\BR_a]e^{-\beta H_a (\BL)}.
\end{equation} 
After performing the path integrals using equations $(3.39-3.41)$ in
\cite{feynman}, we get
\begin{equation}
Z(\{\BR_a\},\{\BR'_a\};L,\BL)=N_0 e^{-\beta \Phi},
\end{equation}
where
\begin{eqnarray}
\Phi &=& A\left(\sum \BR_a^2+\sum\BR_a^{\prime 2}\right)+B\left(\sum 
\BR_a+\sum \BR'_a\right)\cdot \BA 
+2C\sum \BR_a\cdot \BR'_a+ nD\BA^2,
\end{eqnarray}
with
\begin{eqnarray}
A &=& \frac{1}{2} \sqrt{\mu' M} \coth \left( L\sqrt{\mu'/M}
\right)  \\
B &=& \sqrt{\mu' M}
\left[\cosh\left( L\sqrt{\mu'/M }  \right)-1   \right]
\left[\sinh \left( L\sqrt{\mu'/M}\right)\right]^{-1}  
 \\
C &=& -\frac{1}{2} \sqrt{\mu' M}\left[(\sinh \left( L\sqrt{\mu'/M}
\right)\right]^{-1}  \\
D &=& B-L\mu'/2, \\
\BA &=& \frac{2}{\mu'}\sqrt{\sigma}\ \BL .
\end{eqnarray}
The exact form of
the normalization $N_0$ is unimportant as it will cancel out later. 
The next step is to perform the Gaussian integrals over the $\BL$ 
variables in Eq.~(\ref{a1}).  This yields
\begin{eqnarray}
Z_n (\{\BR_a\},\{\BR'_a\};L) &=& N_1 \exp\left\{-U\left(\sum \BR_a^2+
\sum\BR_a^{\prime 2}\right)-
V\left(\sum \BR_a+\sum\BR'_a\right)^2 \right. \nonumber \\
 &-& \left. 2W\sum \BR_a\cdot \BR'_a\right\},
\label{Znf}
\end{eqnarray}
with
\begin{eqnarray}
U &=& \beta A   \\
V &=& -\frac{2\sigma\beta^2 B^2}{{\mu'}^2+8\beta\sigma nD}\\
W &=& \beta C.
\end{eqnarray}
Now, for arbitrary $n$ we can write
\begin{eqnarray}
\langle \BR^2_1(L) \rangle &=&  \frac{\int d\BR_1 \cdots d\BR_n
\BR^2_1 Z_n(\{{\bf 0}\},\{\BR_a\};L)}{\int d\BR_1
\cdots d\BR_n Z_n(\{{\bf 0}\},\{\BR_a\};L)} \nonumber \\
&=&-\frac{1}{n}\frac{d}{dU}\ln \int d\BR_1 \cdots d\BR_n
Z_n(\{{\bf 0}\},\{\BR_a\};L) \nonumber \\
&=& -\frac{1}{n}\frac{d}{2}\frac{d}{dU}  
\ln \left[\frac{\pi^n}{U^{n-1}(U+nV)}\right] \nonumber \\
&=&\ \frac{d}{2} \left( \frac{1}{U} - \frac{V}{U(U+Vn)} \right).
\end{eqnarray}
We have used the fact that the eigenvalues of the matrix associated 
with the quadratic form in Eq. (\ref{Znf}) are $U$ with multiplicity 
$n-1$ and $U+nV$ with multiplicity 1, and the determinant is the 
product of the eigenvalues.
The $n \rightarrow 0$ can now be safely taken to yield
\begin{equation}
\overline{\langle \BR_T^2(L) \rangle}=\lim_{n\rightarrow 0} \langle \BR^2_1
\rangle =\frac{d}{2\beta A}+\frac{\sigma d}{{\mu}^2} 
\left( {\frac{B}{A}} \right) ^2.
\end{equation}
which simplifies to yield Eq.~(\ref{meansq}).

To calculate $\overline{{\langle} {\bf{R}}_T(L){\rangle}^2}$ we use
\begin{eqnarray}
\overline{{\langle} {\bf{R}}_T(L){\rangle}^2}&=& 
\frac{\int d{\bf{R}}_1 
\cdots d{\bf{R}}_n {\bf{R}}_1\cdot{\bf{R}}_2 {Z_n}(\{{\bf{0}}\},\{{\bf{R}}_a\};L)}
{\int d{\bf{R}}_1 \cdots d{\bf{R}}_n 
{Z_n}(\{{\bf{0}}\},\{{\bf{R}}_a\};L)} \nonumber \\
&=&\frac{1}{n(n-1)}\left(\frac{d}{dU}-\frac{d}{dV}\right)
\ln \int d\BR_1 \cdots d\BR_n
Z_n(\{{\bf 0}\},\{\BR_a\};L) \nonumber \\
&=& \frac{1}{n(n-1)}\frac{d}{2}\left(\frac{d}{dU}-\frac{d}{dV}\right)  
\ln \left[\frac{\pi^n}{U^{n-1}(U+nV)}\right] \nonumber \\
&=&\ -\frac{d}{2} \left(\frac{V}{U(U+Vn)} \right).
\label{rt2}
\end{eqnarray}

Next, we show how to calculate $\overline{{\langle} {\bf{R}}_F^2(L){\rangle}}$. 
\begin{eqnarray}
\overline{{\langle} {\bf{R}}_F^2(L){\rangle}}&=& 
\frac{\int \prod d{\bf{R}}_a
\prod d{\bf R}'_a \left({\bf R}_1 - {\bf R}'_1\right)^2 {Z_n}(\{{\bf{R}}_a\},
\{{\bf R}'_a\};L)}{\int \prod d{\bf R}_a \prod d{\bf R'}_a
{Z_n}(\{{\bf{R}}_a\},\{{\bf R}'_a\};L)}. \nonumber \\
&=&\frac{1}{n}\left(\frac{d}{dU}-\frac{d}{dW}\right)
\ln \int \prod d{\bf{R}}_a \prod d{\bf R}'_a
Z_n(\{{\BR_a}\},\{\BR'_a\};L) \nonumber \\
&=& \frac{1}{n}\frac{d}{2}\left(\frac{d}{dU}-\frac{d}{dW}\right)  
\ln \left[\frac{\pi^{2n}}{(U+W)^{n-1}(U-W)^n(U+W+2nV)}\right] \nonumber \\
&=&\ \frac{d}{U-W}=\frac{d}{\beta(A-C)},
\end{eqnarray}
which yields Eq.~(\ref{rsqf}).

Finally, we calculate $\overline{{\langle} {\bf{R}}_Q^2(L){\rangle}}$.
This is given by
\begin{eqnarray}
\overline{{\langle} {\bf{R}}_Q^2(L){\rangle}}&=&
\frac{\int d{\bf{R}}_1 
\cdots d{\bf{R}}_n {\bf{R}}^2_1 {Z_n}(\{{\bf{R}}_a\},\{{\bf{R}}_a\};L)}
{\int d{\bf{R}}_1 \cdots d{\bf{R}}_n 
{Z_n}(\{{\bf{R}}_a\},\{{\bf{R}}_a\};L)} \nonumber \\
&=&-\frac{1}{2n}\frac{d}{dU}\ln \int d\BR_1 \cdots d\BR_n
Z_n(\{\BR_a\},\{\BR_a\};L) \nonumber \\
&=& -\frac{1}{2n}\frac{d}{2}\frac{d}{dU}  
\ln \left[\frac{\pi^n}{[2(U+W)]^{n-1}2(U+W+2nV)}\right] \nonumber \\
&=&\ \frac{d}{4} \left( \frac{1}{U+W} - \frac{2V}{(U+W)(U+W+2nV)} \right).
\end{eqnarray}
In the limit $n \rightarrow 0$ we obtain
\begin{eqnarray}
\overline{{\langle} {\bf{R}}_Q^2(L){\rangle}}&=&
\frac{d}{4} \left( \frac{1}{U+W} - \frac{2V}{(U+W)^2} \right) \nonumber \\
&=& \frac{d\ \sigma}{\mu^2}\left(\frac{B}{A+C}\right)^2+\frac{d}{4\beta(A+C)},
\end{eqnarray}
which gives rise to Eq.~(\ref{shift}). $\overline{{\langle} 
{\bf{R}}_Q(L){\rangle}^2}$ is calculated similarly from $\langle \BR_1 \cdot 
\BR_2 \rangle$ and one obtains $4 \sigma d/\mu^2$ in agreement with 
Eq.~(\ref{shift2}).

\end{document}